\documentclass{elsart3p}
\usepackage{amssymb}
\usepackage{graphicx}
\begin{document}

\begin{frontmatter}
\title{Vortex liquid correlations induced by in-plane field in underdoped Bi$_2$Sr$_2$CaCu$_2$O$_{8+\delta}$}
\author[a]{P. Spathis}
\author[a]{, M. Konczykowski}
\author[a]{, C. J. van der Beek}
\author[b]{, P. Gier{\l}owski}
\author[c]{, M. Li}
\author[c]{, P. H. Kes}
\address[a]{Laboratoire des Solides Irradi\'{e}s, CNRS-UMR7642 \& CEA-DSM-DRECAM, Ecole Polytechnique, 91128 Palaiseau, France}
\address[b]{Institute of Physics of the Polish Academy of Sciences, 32/46 Aleja Lotnik\'ow, 02-668 Warsaw, Poland}
\address[c]{Kamerlingh Onnes Laboratorium, Rijksuniversiteit Leiden, P.O. Box 9506, 2300 RA Leiden, The Netherlands}

\begin{abstract}
By measuring the Josephson Plasma Resonance, we have probed the influence of an in-plane magnetic field on the pancake vortex correlations along the $c$-axis in heavily underdoped Bi$_2$Sr$_2$CaCu$_2$O$_{8+\delta}$ ($T_c=72.4\pm0.6$~K) single crystals both in the vortex liquid and in the vortex solid phase. Whereas the in-plane field enhances the interlayer phase coherence in the liquid state close to the melting line, it slightly depresses it in the solid state. This is interpreted as the result of an attractive force between pancake vortices and Josephson vortices, apparently also present in the vortex liquid state. The results unveil a boundary between a correlated vortex liquid in which pancakes adapt to Josephson vortices, and the usual homogeneous liquid.
\end{abstract}

\begin{keyword}
Josephson Plasma Resonance\sep Bi:2212 \sep Josephson vortex
\PACS 74.25.Qt \sep 74.50.+r
\end{keyword}
\end{frontmatter}

\section{Introduction}
\label{intro}
In highly anisotropic layered superconductors, tilting a magnetic field at a direction oblique to the CuO$_2$ planes leads to novel states of vortex matter. Bitter decoration\cite{bolle}, Hall probe microscopy and Lorentz microscopy have confirmed that the attraction between Josephson vortex (JV) stacks and pancake vortices (PV) in the vortex solid state of Bi$_2$Sr$_2$CaCu$_2$O$_{8+\delta}$ leads to the so-called crossing-lattice\cite{koshelevCROSSING}. This state is only one of many constituting a particularly rich phase diagram. The occurence of other structural phases of vortex matter, \textit{e.g.}, the lattice of tilted PV stacks, the combined perpendicular and tilted PV stacks\cite{konc}, or the PV-soliton lattice, depends on the interplay between the magnetic and Josephson coupling contributions to the PV lattice tilt modulus \cite{koshelev2005}.

The Josephson Plasma Resonance (JPR) frequency $f_{JPR}$ is sensitive to the superconducting phase difference, $\Phi_{n,n+1}$, between CuO$_2$ layers $n,n+1$, through $f_{JPR}^2(B_\perp,B_\parallel,T)=f_{JPR}^2(0,0,T) \langle \cos \Phi_{n,n+1} \rangle$ where $\langle \ldots \rangle$ stands for thermal and disorder average. Since $\Phi_{n,n+1}$ intimately depends on the alignment of PV stacks, JPR can in principle be used to detect and identify different vortex phases.
\section{Experiment}
\label{exp}
The Bi$_2$Sr$_2$CaCu$_2$O$_{8+\delta}$ single crystals ($T_c=72.4\pm0.6$~K) were cut from a larger underdoped Bi$_2$Sr$_2$CaCu$_2$O$_{8+\delta}$ crystal, grown by the travelling solvent floating zone technique. The JPR was measured using the cavity perturbation technique in the TM$_{01i}$ modes, with the microwave electrical field aligned along the sample $c$-axis. Two orthogonal coils were used to apply field components parallel ($H_\parallel$) and perpendicular ($H_\perp$) to the CuO$_2$ layers. Varying the mode $i$, allowed us to change the JPR frequency and to probe both the vortex solid and the liquid state.
\section{Results and discussion}
\begin{figure}[t]
\begin{center}
\includegraphics[width=\columnwidth,keepaspectratio]{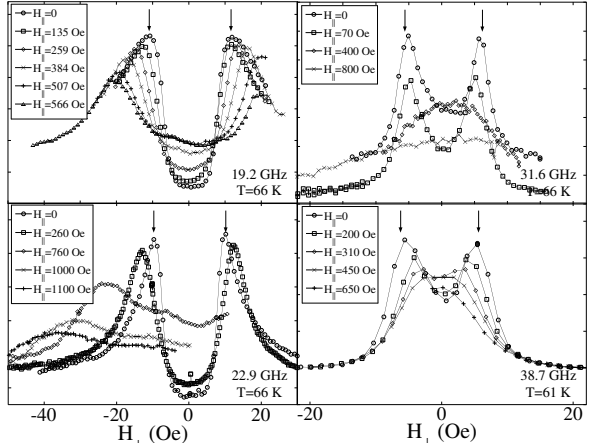}
\caption{\label{fig:dissipation}Magnetic-field dependance of the microwave absorption (arbitrary units) of a Bi$_2$Sr$_2$CaCu$_2$O$_{8+\delta}$ single crystal obtained at different frequencies.}
\end{center}
\end{figure}
Figure \ref{fig:dissipation} shows the microwave absorption obtained by sweeping $H_\perp$ at constant $H_\parallel$ and temperature. When $H_\parallel=0$, the JPR is identified as the maximum of the microwave absorption, for the field $H_\perp^{JPR}$ (arrows in figure \ref{fig:dissipation}). Following the evolution of the absorption, the lineshape is modified by $H_\parallel$ in two ways. First, the intensity of the microwave response decreases for all frequencies. It has been pointed out that the presence of a Josephson vortex lattice (JVL) strongly modulates the \textit{c}-axis critical current\cite{bulaevskii1996}. Thus, the JPR cannot develop in the stacks of JVs. However, for low values of the in-plane field ($H_\parallel\ll \Phi_0/\gamma s^2$, where $\gamma$ is the anisotropy ratio and $s$ the interlayer distance), the distance between two stacks of JVs is sufficiently high so that the phase remains unaffected far from the cores. The remaining microwave absorption mainly arises from JV-free regions, the extent of which decreases linearly with $H_\parallel$.

Second, $H_\parallel$ changes the field $H_\perp^{JPR}$ at which the maximum absorption occurs: the decrease of $H_\perp^{JPR}$ observed in the vortex solid at 31.2 and 39.4 GHz implies the decrease of $\langle \cos \Phi_{n,n+1} \rangle$ and is consistent with the addition of a JVL. However, for 19.2 and 22.9 GHz, $H_\perp^{JPR}$, measured in the vortex liquid, \textit{increases}.

\begin{figure}[t]
\begin{center}
\includegraphics[width=\columnwidth,keepaspectratio]{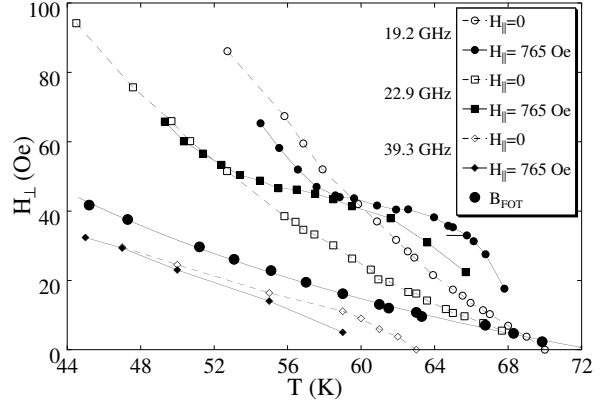}
\caption{\label{fig:result}Temperature dependance of $H_\perp^{JPR}$ obtained for various in-plane fields. The melting line $B_{FOT}$ with no in-plane field is also reported.}
\end{center}
\end{figure}

The $H_\perp^{JPR}$-loci for different temperatures are shown in Fig. \ref{fig:result} together with the PV vortex lattice melting line in $H_\parallel=0$. The increase of $H_\perp^{JPR}$ is only observed in the vortex liquid state and for sufficiently low PV densities. In the solid phase, it has been shown\cite{tsuiJPR1,matsuda} that the addition of PVs on a dense lattice of JVs ($H_\parallel\gg\Phi_0/\gamma s^2$) increases the $c$-axis critical current: the PVs adjust their positions to the JVL so as to maximize $\langle \cos \Phi_{n,n+1} \rangle$\cite{bulaevskii1996}. The same effect could be present in the vortex liquid phase, giving rise to an effective attraction between PVs and JVs and to a \textit{correlated vortex liquid} in which the density of PVs is smaller in JV-free regions. Since the microwave absorption due to the JPR mainly comes from those regions, increasing $H_\parallel$ also increases $H_\perp^{JPR}$. However, even though the presence of well-defined Josephson vortices in the presence of a PV liquid is still controversial, we believe that it can be the case for low densities of PVs. The correlated vortex liquid is therefore stable close to the PV melting line, for low values of the perpendicular magnetic field.

\end{document}